\def\year{2017}\relax
\documentclass[letterpaper]{article}
\usepackage{aaai17}
\usepackage{times}
\usepackage{helvet}
\usepackage{courier}
\usepackage{amsmath,amsfonts,amssymb,bbm,amsthm} 
\usepackage{algorithm}  
\usepackage{algorithmic}
\usepackage{multirow}
\usepackage{extarrows}
\usepackage{graphicx}
\graphicspath{ {images/} }

\newtheorem{theorem}{Theorem}
\newtheorem{lemma}{Lemma}
\newtheorem{proposition}{Proposition}

\newtheorem{definition}{Definition}
\newtheorem{remark}{Remark}

\begin{document}
\title{Non-additive Security Games}
\author{Sinong Wang, Fang Liu, Ness 	Shroff\\
The Ohio State University, 2015 Neil Avenue, OH 43210, USA\\
\{wang.7691, liu.3977, shroff.11\}@osu.edu
}
\maketitle
\begin{abstract}
We have investigated the security game under non-additive utility functions.
\end{abstract}

\section{Introduction}

The nature of resource allocation in practical \emph{security games} often results in exponentially many pure strategies for the defender, such that the defender's optimal mixed strategy is hard to solve. In the past few years, several works have tried to resolve this issue from both theoretical and practical perspectives~\cite{kiekintveld2009computing,korzhyk2010complexity,jain2011double,letchford2013solving,xu2014solving,xu2016mysteries}. A common restriction in these works is to assume that the attacker only attacks one target or that different targets are independent. The latter implies that the payoff of a group of targets is the sum of the payoffs of each one~\cite{korzhyk2011security}. In practice, there exists various linkage structures among the targets such that attacking one target will influence the others. Traditional models that ignore the inherent synergy effect between the targets could lead to catastrophic consequences~\cite{buldyrev2010catastrophic}. Motivated by this phenomenon, there are some recent works that investigate the security game with dependent targets~\cite{shakarian2014power,vorobeychik2015securing}.

However, these works are limited to specific dependencies and do not provide a systematic understanding of complexity properties or provide an efficient algorithm. For example, Shakarian et al.~\shortcite{shakarian2014power} assumes that the attacker and defender can choose a subset of all nodes in a power gird and their utilities are dependent on the set of disconnected loads. They show that the defender best response problem (DBR) can be solved in polynomial time if the attacker attacks at most one target, while NP-hard in other cases. However, their complexity results cannot be easily reduced to the complexity of determining defender's mixed strategies. 

In this paper, we introduce a new security game, we call the Non-additivity Security Game (NASG), which is a non-zero-sum game including two players - the \emph{defender} and \emph{attacker}, and $n$ targets, denoted by $[n]=\{1,2,\ldots,n\}$. We model various dependencies of targets by defining the strategy of each player as a subset of $[n]$ and adopt a general set function as the utilities. Specifically, the attacker will obtain benefits for successfully attacked targets and pay a cost for its strategy. Also, the defender will lose benefits for those targets and also pays a cost. A critical point in the NASG is that \emph{the benefit and cost for several targets is not the summation of each target's utility, instead, it is dependent on the specific combination of targets.}.  

At a high level, the main challenge of NASG is that both the size of the strategy space and the number of utility functions are $\Theta(2^n)$. We wonder how the following questions that are well understood in the case of additive utility functions can be addressed in the case of non-additivity assumption.
\begin{itemize}
	\item How to \textbf{compactly} represent the NASG and how to \textbf{efficiently compute} the mixed strategies of NASG?
	\item What is the \textbf{complexity} of computing the mixed strategies of NASG?
\end{itemize}

To answer these questions, we make the following contributions: (1) We provide the condition for compactly representing NASG and prove that there exists $\text{poly}(n)$ number of variables in the compact model if the number of non-additive utility functions is $\text{poly}(n)$. The main technique is isomorphism and projection of a polytope. (2) We design an algorithmic framework to efficiently compute the mixed strategies for NASGs by reducing the original problem to an oracle problem. The main technique is to design a polynomial-time vertex mapping algorithm from the low-dimensional polytope to a simplex; (3) We prove that above oracle problem and the computation of mixed strategies of NASG can be reduced to each other in polynomial-time under a reasonable restriction. Furthermore, we show that such an oracle problem is a problem of maximizing a \emph{pseudo-boolean function}; (4) Finally, we apply our theoretical framework to the \emph{network security game}. We provide polynomial-time algorithms for some kinds of networks and security measures, while for the general case, we show the NP-hardness and propose an approximation algorithm. 

All the proofs in this paper are left to the supplemental material due to the space limitation.

\section{Problem Description and Preliminary}

We begin by defining the NASG as a two-player normal-form non-zero-sum game.

\textbf{Players} and \textbf{targets:} The NASG contains two players (a \emph{defender} and an \emph{attacker}), and $n$ targets, indexed by set $[n]\triangleq \{1,2,\ldots,n\}$.

\textbf{Strategies and utility functions:} A \emph{pure} strategy for each player is a subset of $[n]$. In the general case, we consider the \emph{complete pure strategy space} of attacker and defender, defined as the power set $2^{[n]}\triangleq\{V|V\subseteq[n]\}$, denoted by $\mathcal{A}$ and $\mathcal{D}$, respectively. So there are $N\triangleq 2^n$ pure strategies for both players. Let \textbf{set function} $C_{a}(\cdot): \mathcal{A}\rightarrow \mathbb{R}$ and $C_{d}(\cdot): \mathcal{D}\rightarrow \mathbb{R}$ be the attacker's and defender's cost function, and the set function $B(\cdot): \mathcal{A}\rightarrow \mathbb{R}$ be the benefit function. 

\begin{remark}
Traditional models do not consider a cost function, instead, they assume there exists a resource constraint such that certain strategies, i.e., subsets of $[n]$, are restricted. In our paper, we explicitly consider the cost function but do not have such resource constraints\footnote{Later, in this paper, we will consider the limited resource.}. In cybersecurity applications, security resources are available for a cost and can be used to replace resource constraints, as illustrated in~\cite{vorobeychik2015securing}.
\end{remark}

\textbf{Tie-breaking Rule: }When the attacker and defender choose strategy $A\in\mathcal{A}$ and $D\in\mathcal{D}$, targets in the set $A\backslash D$ are successfully attacked by the attacker. Moreover, both players should pay the cost for their strategy, and the attacker's and defender's payoff is given by $[B(A\backslash D)-C_{a}(A)]$ and $[-B(A\backslash D)-C_{d}(D)]$, respectively. 

\textbf{Normal-form representation:} Suppose that the order of the attacker's pure strategy is given by index function $\sigma(\cdot): \mathcal{A}\rightarrow \{1,2,\cdots,N\}$, then define the index function $\mu(\cdot)$ for defender's pure strategy\footnote{This definition of the index function is to guarantee the symmetry of benefit matrix, which simplifies most theoretical results.}: $\mu(U)=\sigma(U^c)$ for any $U\in\mathcal{D}$. Then we can define the utility matrices including the cost matrices of attacker and defender: $\mathbf{C}^A,\mathbf{C}^D\in\mathbb{R}^{N\times N}$, 
\begin{equation*}
	\mathbf{C}^A_{\sigma(A),\mu(D)}=C_a(A), \mathbf{C}^D_{\sigma(A),\mu(D)}=C_d(D),\forall A, D\in 2^{[n]},
\end{equation*}
and the benefit matrix $\mathbf{M}\in\mathbb{R}^{N\times N}$,
\begin{equation*}
	\mathbf{M}_{\sigma(A),\mu(D)}=B(A\backslash D), \forall A, D\in 2^{[n]}.
\end{equation*}
Let $\mathbf{M^{a}}$ and $\mathbf{M^{d}}$ be the attacker's and defender's payoff matrices. It's clear that $\mathbf{M^a}=\mathbf{M}-\mathbf{C}^A$ and $\mathbf{M^d}=-\mathbf{M}-\mathbf{C}^D$. The \emph{mixed} strategy $\mathbf{p}, \mathbf{q}\in \Delta_N$ is a distribution over the set of pure strategy $\mathcal{A},\mathcal{D}$, where $\mathbf{p}_{\sigma(A)}, \mathbf{q}_{\mu(D)}$ is the probability that attacker choose strategy $A$ and defender choose strategy $D$. ${\Delta}_N$ represents a $N$-dimensional simplex. Then the expected payoffs for the attacker and defender is given by following bilinear form, when they play the mixed strategy $\mathbf{p}\in\Delta_N$ and $\mathbf{q}\in\Delta_N$, by 
\begin{align*}
U_a(\mathbf{p},\mathbf{q})=\mathbf{p}^T\mathbf{M^aq} \quad\text{and}\quad U_d(\mathbf{p},\mathbf{q})=\mathbf{p}^T\mathbf{M^dq}.
\end{align*}

\begin{figure}[t]
\vspace{-0.1in}
\leftline{\includegraphics[width=0.46\textwidth]{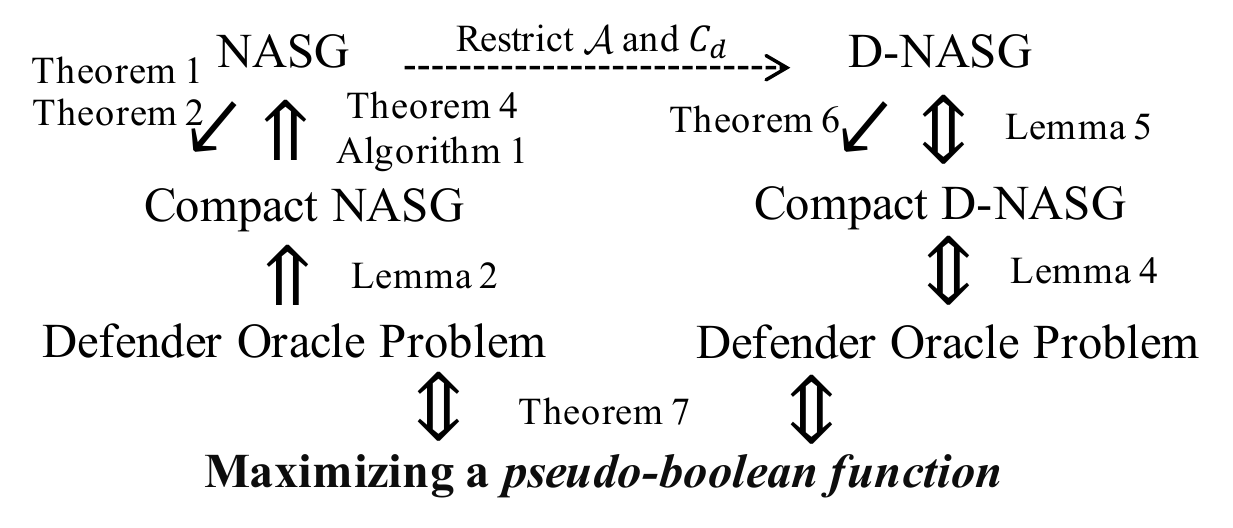}}
\vspace{-.2in}
\caption{The summary of main results. The double arrow denotes the polynomial time reduction. The single arrow denotes the compact representation}
\vspace{-0.2in}
\label{fig:lowrank}
\end{figure}

\textbf{Solution Concepts:} In this paper, we assume that both players move simultaneously and the standard solution concept is the \emph{Nash Equilibrium (NE)}. Our goal is to compute the defender's minimax mixed strategies and we call it the \textbf{min max problem}.

The following three definitions are heavily used in our theoretical development.
\begin{definition}\label{def:com}
 	The common utility is defined as the transform of the benefit function $B(\cdot)$ for all $U\in 2^{[n]}$,
	\begin{align*}
		&B^c(U)=\sum\limits_{V\subseteq U}(-1)^{|U\backslash V|}B(V).
	\end{align*}
\end{definition}
Similar definitions hold for cost functions: $C_a(\cdot)$, $C_{d}(\cdot)$ and their common utility: $C^c_a(\cdot)$, $C^{c}_d(\cdot)$, and the details can be seen in the supplemental material. 
\begin{definition}
The support set of NASG is
\begin{equation}\label{eq:supportset}
	S=\{U\in \mathcal{A}|B^c(U)\text{ or }C^c_a(U)\text{ or }C^c_d(U^c)\neq 0\},
\end{equation}
and support index set $\sigma(S)=\{\sigma(U)|U\in S\}$.
\end{definition}
\begin{definition}
The projection operator $\pi_S: \mathbb{R}^N\rightarrow\mathbb{R}^{|S|}$ is
\begin{equation}
	\pi_S((\mathbf{x}_1,\mathbf{x}_2,\ldots,\mathbf{x}_N))=(\ldots,\mathbf{x}_{i},\ldots)_{i\in \sigma(S)},
\end{equation} 
and projection of polytope: $\Pi_{S}(\Delta_N)\triangleq \{\pi_S(\mathbf{x})|\mathbf{x}\in\Delta_N\}$. 
\end{definition}

\subsection{Strategically Zero-sum Form}
Although a NASG contains non-zero-sum payoff, we prove the following proposition, which shows that it belongs to the strategically zero-sum game~\cite{moulin1978strategically}.
\begin{proposition}\label{prop:zerosum}
The set of Nash equilibriums of NASG is equivalent to the set of Nash equilibriums of zero-sum game with payoff matrix $\mathbf{M}-\mathbf{C^{A}}+\mathbf{C^{D}}$. 
\end{proposition}
Clearly, the stackelberg equilibrium set is equivalent to the NE set of the NASG. This proposition allows us to solve the NASG via the equivalent zero-sum game, which can be tackled by a linear programming approach. In the sequel, we use $\mathbf{M}^{\circ}=\mathbf{M}-\mathbf{C^{A}}+\mathbf{C^{D}}$ to denote the payoff matrix.
\begin{remark}
Traditional zero-sum security game~\cite{xu2016mysteries} assumes that the defender gets a reward $r_i$ if target $i$ is covered, or incurs a cost $c_i$ if uncovered. We can set specific values for our utility functions to recover their setting.
\end{remark}
The main results of this paper is summarized in Fig.~\ref{fig:lowrank}.

\section{The Compact Representation of NASG}

Based on the equivalent zero-sum game $\mathbf{M}^{\circ}$ and von Neumann's minimax theorem, computing the NE of NASG can be formulated as the following min max problem,
\begin{align}
	&\min\limits_{\mathbf{q}\in \Delta^N}\max\limits_{\mathbf{p}\in\Delta^N}\mathbf{p}^T\mathbf{M}^{\circ}\mathbf{q}\label{eq:orgmodel}.
\end{align} 
This optimization model has $2^{n+1}$ variables, which implies that NASG is in general hard to solve. The goal of this section is to find a condition on the NASG that can be compactly represented with only $\text{poly}(n)$ variables. To convey our idea more easily, we begin with an example.

\subsection{Motivating Example}

We first conduct the gauss elimination of the matrix $\mathbf{M}^{\circ}$ to transform it into the row canonical form, which is to left and right multiply $\mathbf{M}^{\circ}$ by elementary matrices $\mathbf{E}$ and $\mathbf{F}$,
\begin{align}
	\min\limits_{\mathbf{q}\in\Delta_{N}}\max\limits_{\mathbf{p}\in\Delta_{N}}\mathbf{p}^T\mathbf{M}^{\circ}&\mathbf{q}=\min\limits_{\mathbf{q}\in\Delta_{N}}\max\limits_{\mathbf{p}\in\Delta_{N}}\mathbf{p}^T\mathbf{E}^{-1}\mathbf{E}\mathbf{M}^{\circ}\mathbf{F}\mathbf{F}^{-1}\mathbf{q}\notag\\
	&=\min\limits_{\mathbf{q}\in\Delta_{N}}\max\limits_{\mathbf{p}\in\Delta_{N}}\mathbf{p}^T\mathbf{E}^{-1}\begin{bmatrix}
\mathbf{M}^{\circ}_{r} & \mathbf{0}\\ 
\mathbf{0} & \mathbf{0}
\end{bmatrix}\mathbf{F}^{-1}\mathbf{q}.\notag
\end{align}
where $r$ is the rank of payoff matrix $\mathbf{M}^{\circ}$ and $\mathbf{M}^{\circ}_{r}$ is the non-zero block of its row canonical form. If we define the affine projection $f(\mathbf{p})=\left(\mathbf{p}^T\mathbf{E}^{-1}\right)^T, g(\mathbf{q})=\mathbf{F}^{-1}\mathbf{q}$, and let $\Delta^{a}_{N}=\{f(\mathbf{p})| \mathbf{p} \in \Delta_{N}\}$, $\Delta^{d}_{N}=\{g(\mathbf{q})|\mathbf{q} \in \Delta_{N}\}$, we can obtain the following optimization problem,
\begin{align}
	&\min\limits_{\mathbf{q}'\in\Delta^{d}_{N}}\max\limits_{\mathbf{p}'\in\Delta^{a}_{N}}\mathbf{p'}^T\begin{bmatrix}
\mathbf{M}^{\circ}_{r} & \mathbf{0}\label{eq:affmodel}\\ 
\mathbf{0} & \mathbf{0}
\end{bmatrix}\mathbf{q}'.
\end{align}

Since the polyhedra $\Delta^{a}_{N}$ and $\Delta_{N}$ are isomorphic, their vertices exhibit the one-one correspondence. Similar argument for the polyhedra $\Delta^{d}_N$ and $\Delta_{N}$. Thus the optimization problem (\ref{eq:orgmodel}) and (\ref{eq:affmodel}) is equivalent. Further, considering the fact that only the first $r$ elements in vector $\mathbf{p'}$ and $\mathbf{q'}$ have the non-zero coefficients in (\ref{eq:affmodel}), we can further simplify it as
\begin{align}
	&\min\limits_{\mathbf{q'}\in\Pi_{r}\left(\Delta^{d}_{N}\right)}\max\limits_{\mathbf{p'}\in\Pi_{r}\left(\Delta^{a}_{N}\right)}\mathbf{p'}^T\mathbf{M}^{\circ}_{r}\mathbf{q'}\label{eq:excptmodel},
\end{align}
where the operator $\Pi_{r}(\cdot)$ is to project the $N-$dimensional polytope into its first $r-$coordinates.
\begin{remark}
	The observation is that the number of variables in the model (\ref{eq:excptmodel}) depends on the rank of payoff matrix. For example, if the rank of $\mathbf{M}^{\circ}$ is \emph{poly}($n$), we can compactly represent NASG with only \emph{poly}($n$) variables.
\end{remark}

\subsection{The Formal Description of Compact NASG}

Although the above conceptual derivation provides a possible path to compactly represent the NASG, there exists a significant technical challenge: the elementary matrices $\mathbf{E}$, $\mathbf{F}$ and their inverse matrices have exponential size, can we \textbf{find both elementary matrices} efficiently? To tackle this problem, we first show that the payoff matrix $\mathbf{M}^{\circ}$ can be decomposed as the product of three matrices. The following technical lemma is critical in our decomposition.
\begin{lemma}\label{lm:com}
For all $U\in 2^{[n]}$, benefit function satisfies
	\begin{align*}
		&B(U)=\sum\limits_{V\subseteq U}B^c(V).
	\end{align*}
\end{lemma}
Note that similar results hold for the cost functions and their common utility. Then we have the decomposition of payoff matrix $\mathbf{M}^{\circ}$ in terms of common utilities and an illustrative example can be seen in the supplemental material.
\begin{theorem}\label{thm:decomp}
The payoff matrix $\mathbf{M}^\circ=\mathbf{M}-\mathbf{C}^A+\mathbf{C}^D$ can be decomposed as 
	\begin{equation}
	\mathbf{M}^\circ=\mathbf{Q}(\mathbf{D}-\mathbf{L}+\mathbf{V})\mathbf{Q}^T,
	\end{equation}	
	where $\mathbf{D}$ is the diagonal matrix with $\mathbf{D}_{\sigma(A),\sigma(A)}=B^c(A)$. $\mathbf{V}$  and $\mathbf{L}$ are two sparse matrices with non-zero elements: $\mathbf{V}_{\mu([n]), \sigma(A)}=C^c_d(A), \mathbf{L}_{\mu(D),\sigma(\{\emptyset\})}=C_a^c(D^c)$. The $\mathbf{Q}$ is binary matrix with $\mathbf{Q}_{\sigma(A),\mu(D)}=$ $\mathbbm{1}\{D^c \subseteq A\}$.
\end{theorem}
Based on this result, we can let elementary matrices $\mathbf{E}=\mathbf{Q}^{-1}$, $\mathbf{F}=(\mathbf{Q}^T)^{-1}$, affine transformation $f(\mathbf{p})=\mathbf{Q}^T\mathbf{p}$ and $g(\mathbf{q})=\mathbf{Q}^T\mathbf{q}$ to yield two isomorphic polytopes: $\Delta^{a}_N$ and $\Delta^{d}_N$ with $\Delta_N$. The whole procedure is listed in Fig.~\ref{fig:framework}, and the following theorem answers part of our first question.

\begin{figure}[t]
\normalsize
\vspace{-0.1in} 
\begin{tabular}{|p{8cm}|}
\hline
\vspace{0.01in} 
\textbf{$\qquad$The Framework of Compact Representation}\\
\vspace{0.01in} 
\textbf{Isomorphic polytope:}  solving the NASG is equivalent to solving the following optimization problem,
\begin{equation}\label{eq:frame1}
	\min\limits_{\mathbf{q'}\in \Delta^{d}_N}\max\limits_{\mathbf{p'}\in\Delta^{a}_N}\mathbf{p'}^T(\mathbf{D}-\mathbf{L}+\mathbf{V})\mathbf{q'}
\end{equation}
\\
\textbf{Projection of polytope:} projects the polytope $\Delta^{a}_N$ and $\Delta^{d}_N$ into coordinates with indices in $\sigma(S)$, and further simplify (\ref{eq:frame1}) as the following compact represented model,
\begin{equation}
	\textbf{Compact NASG}\min\limits_{\mathbf{q'}\in\Pi_{S}\left(\Delta^{d}_{N}\right)}\max\limits_{\mathbf{p'}\in\Pi_{S}\left(\Delta^{a}_{N}\right)} \mathbf{p'}^T\mathbf{M}^S\mathbf{q'}\label{thm:eq:cptmodel}
\end{equation}
where matrix $\mathbf{M}^S$ is a sub-matrix of $\mathbf{D}-\mathbf{L}+\mathbf{V}$, which is obtained by extracting those rows and columns whose index belonging to $\sigma(S)$. \\
\hline
\end{tabular}
\vspace{-0.1in}
\caption{{The isomorphism and projection of a polytope}}
\label{fig:framework}
\vspace{-0.25in}
\end{figure}

\begin{theorem}	\label{thm:cpt}
If $|S|=\text{\emph{poly}}(n)$, the rank of the payoff matrix $\mathbf{M}^\circ$ is \emph{poly}$(n)$, moreover, the NASG can be compactly represented by \emph{poly}$(n)$ number of variables and $(\mathbf{p}^*,\mathbf{q}^*)\text{ is a NE of NASG }$ if and only if $(\pi_S(f(\mathbf{p}^*)),$ $\pi_{S}(g(\mathbf{q}^*)))$ is the optimal solution of (\ref{thm:eq:cptmodel}).
\end{theorem}
Since we do not utilize the row canonical form of $\mathbf{M}^{\circ}$, instead, we extract the non-zero columns and rows of $\mathbf{D}-\mathbf{L}+\mathbf{V}$ to form the low-dimensional matrix $\mathbf{M}^{S}$, the Theorem~\ref{thm:cpt} provides only a sufficient condition for our compact representation. Indeed, we can make it both sufficient and necessary by further conducting elementary elimination to transform the matrix $\mathbf{D}-\mathbf{L}+\mathbf{V}$ into an approximate diagonal matrix $\mathbf{D}$. However, this process will significantly complicate our affine transformation $f$ and $g$, and make it impossible to map the optimal solution of compact model (\ref{thm:eq:cptmodel}) to the original mixed strategy.

\textbf{Implication of compact NASG:} From the perspective of attacker's utility function, our compact representation (\ref{thm:eq:cptmodel}) simplifies $U_a(\mathbf{p},\mathbf{q})$ as (similar result holds for $U_d(\mathbf{p},\mathbf{q})$),
\begin{equation}\label{eq:explain}
	\sum_{U\in S} \mathbf{p}_{\sigma(U)}'[\mathbf{q}_{\sigma(U)}' B^c(U)-C^c_a(U)].
\end{equation}
Based on the definition of affine transformation $f,g$ and matrix $\mathbf{Q}$, each variable $\mathbf{p}_{\sigma(U)}'=\sum_{V:U\subseteq V}\mathbf{p}_{\sigma(V)}$ is the probability that the attacker attacks all the targets in set $U$, while $\mathbf{q}_{\sigma(U)}'=\sum_{V\subseteq U^c}\mathbf{q}_{\mu(V)}$ is the probability that defender does not defend any any targets in set $U$. Therefore, we can regard $B^c(U)$, $C^c_a(U)$ (not $B(U)$ and $C_a(U)$!) as the benefit and cost function for a ``virtual target $U$'', and the formula (\ref{eq:explain}) calculates the expected utility in such a new game. 
\begin{remark}
	If all the utility functions are additive, then all the common utility functions are zero except those defined on the singleton set. Thus $S=[n]\cup\{\emptyset\}$ and each variable $\mathbf{p}_{\sigma(\{i\})}'$, $\mathbf{q}_{\sigma(\{i\})}'$ in the compact model is a marginal probability that target $i$ is attacked or defended, thus our framework recovers the results in~\cite{kiekintveld2009computing}. 
	%Another observation is that non-additivity of strategy $U$ is not described by the non-zero difference of its additive counterpart, i.e., $B(U)-\sum_{i\subseteq U}B(i)$, instead, by a specific parameter, common utility $B^c(U), C^c_a(U)$ and $C^c_d(U)$.
\end{remark}

In the sequel, \emph{the terminology NASG refers to those NASGs those only have $\text{poly}(n)$ variables in their compact model}.  Based on our compact representation, a natural question arising is that, can we efficiently solve such compact model and implement the optimal solution by the defender's mixed strategy? We will answer this question in the next section. 

\section{Oracle-based Algorithmic Framework}
The main result of this section is given in the following theorem. 
\begin{theorem}
There is a \emph{poly}$(n)$ time algorithm to solve the min max problem, if there is a \emph{poly}$(n)$ time algorithm to compute the defender oracle problem, defined as, for any vector $\mathbf{w}\in\mathbb{R}^{|S|}$, compute
	\begin{equation}\label{eq:dop}
		\mathbf{x}^*=\arg\max\limits_{\mathbf{x}\in \Pi_{S}(\Delta^{d}_{N})}\mathbf{w}^T\mathbf{x}.
	\end{equation}
\end{theorem}

It is not surprising that the compact NASG can be reduced to the defender oracle problem (DOP), and the reduction follows from an application of equivalence between separation and optimization~\cite{grotschel1981ellipsoid}. What is interesting, however, is the reduction from the \textbf{min max problem} to the \textbf{compact NASG}. Namely, how to map the optimal solution of compact NASG to a defender's mixed strategy in $\text{poly}(n)$ time. We show that this could be done by exploiting the geometric structure of polytope $\Delta^{d}_{N}$.

\subsection{Reducing Compact NASG to Oracle Problem}

For simplicity, we use $H_a$ and $H_d$ to denote the polytope $\Pi_{S}(\Delta^{a}_{N})$ and $\Pi_{S}(\Delta^{d}_{N})$, and $I_a$ and $I_d$ to denote their vertices, respectively. The compactly represented NASG (\ref{thm:eq:cptmodel}) can be formulated as the linear programing (LP) problem.
\begin{align}
\textbf{Compact-LP}\quad&\min\quad u\label{eq:compact}\\
& \begin{array}{r@{\quad}l@{}l@{\quad}l}
s.t. &\mathbf{v}^T\mathbf{M}^S\mathbf{q}'\leq u\quad\forall\mathbf{v}\in I_a, \label{eq:payoffconstraint}\\
 &\mathbf{q}'\in H_d. 
\end{array} 
\end{align}

The compact LP has $\text{poly}(n)$ number of variables and possibly exponentially many constraints. One can therefore apply the ellipsoid method to solve such an LP, given a $\text{poly}(n)$ time separation oracle. Further, the separation oracle can be reduced to the following two parts: given any $(\mathbf{q}',u)$, (1) \emph{membership problem}: decide whether $\mathbf{q}'\in H^d$. If not, generate a hyperplane that separating $(\mathbf{q}'$, $u)$ and $H^d$; (2) \emph{inequality constraint problem}: decide whether all the inequality constraints hold. If not, find one violating constraint. We have the following result for these problems.  
\begin{lemma}\label{lm:miop}
	The \emph{membership problem} and \emph{inequality constraint problem} of compact LP (\ref{eq:compact}) can be reduced to the defender oracle problem (\ref{eq:dop}) in \emph{poly}$(n)$ time.
\end{lemma}

\subsection{Reducing NASG to Compact NASG}

A classical result in combinatorial optimization is that if the separation problem of polytope $P\in\mathbb{R}^n$ can be solved in $\text{poly}(n)$ time, we can decompose any point $\mathbf{x}\in P$ into the convex combination of at most $(n+1)$ vertices of $P$~\cite{grotschel1981ellipsoid}. Note that this is precisely the DOP required for above reduction. Applying this result to the optimal solution $\mathbf{x}^*$ of compact LP (\ref{eq:compact}), we can get a convex decomposition that $\mathbf{x}^*=\sum_{i=1}^{|S|+1}\lambda_i \mathbf{v}^i$, where $\mathbf{v}^i\in I_d$. If we can map the vertices $\mathbf{v}^i$ back to the vertices (pure strategy) of original NASG, denoted by $h(\mathbf{v}^i)$, the mixed strategies of defender can be expressed as 
\begin{equation}
	\mathbf{q}^*=\sum_{i=1}^{|S|+1}\lambda_i h(\mathbf{v}^i).
\end{equation}
Thus, the key lies in how to compute $h(\mathbf{v}^i)$ in $\text{poly}(n)$ time. 

To tackle this problem, first, considering an arbitrary pure strategy $U\in 2^{[n]}$, the corresponding vertex is a unit vector $\mathbf{e}^U\in \mathbb{R}^N$ with only one non-zero element $\mathbf{e}^U_{\mu(U)}=1$. Based on the affine transformation $g(\mathbf{q})=\mathbf{Q}^T\mathbf{q}$, the corresponding vertex of isomorphic polytope $\Delta^d_N$ is
\begin{equation}
	g(\mathbf{e}^U)=\mathbf{Q}^T\mathbf{e}^U=\mathbf{Q}_{\mu(U)}^T,
\end{equation}
where $\mathbf{Q}_{\mu(U)}$ is the $\mu(U)$th row of matrix $\mathbf{Q}$. Then the corresponding point $\mathbf{v}^U$ of the projected polytope $H^d$ is 
\begin{equation}
	\mathbf{v}^U=\pi_{S}(\mathbf{Q}_{\mu(U)}^T),
\end{equation}
which is a sub-vector of $\mathbf{Q}_{\mu(U)}^T$. The problem is that the vertex in the high-dimensional polytope may not project to a vertex of its low-dimensional image. However, the following lemma will provide a positive result.
\begin{lemma}\label{lm:cptvertex}
$\forall S\subseteq 2^{[n]}$ s.t. $[n]\subseteq S$, the vertices of the polytope $H_d$ are the rows of a sub-matrix of $\mathbf{Q}$, which is formed by extracting the column whose index belongs to $\sigma(S)$. 
\end{lemma}

No matter which coordinate we project the polytope $\Delta^d_N$ into, the number of vertices is still $N$, and they forms a sub-matrix of $\mathbf{Q}$. Therefore, we can exploit the property of matrix $\mathbf{Q}$ to construct a vertex mapping algorithm and the correctness of Algorithm~\ref{alg1} is justified by following theorem.

\begin{algorithm}[t]
\renewcommand{\algorithmicrequire}{ \textbf{Input:}} %Use Input in the format of Algorithm  
\renewcommand{\algorithmicensure}{ \textbf{Output:}} %UseOutput in the format of Algorithm 
\caption{Vertex Mapping from Vertex to Pure Strategy}   
\label{alg1}  
\begin{algorithmic}  
\REQUIRE Vertex $\mathbf{v}^U\in I^d$  
\ENSURE Defender's pure strategy $U$ of original NASG
\STATE $T=\emptyset$;
\FOR{each $i \in [n]$} 
\STATE \textbf{if} $\mathbf{v}^U_{\sigma(\{i\})}\neq 0$ \textbf{ then} $T=T\cup\{i\}$;
\ENDFOR
\STATE $U=T^c$;
\end{algorithmic}  
\end{algorithm}

\begin{theorem}\label{thm:cpt2org}
	Vertex mapping algorithm runs in $O(n)$ time and maps each vertex of $H_d$ to a unique pure strategy.
\end{theorem}

\section{Solving NASG is a Combinatorial Problem}

In this section, we will answer our second question, i.e., what is the complexity of the NASG, in a restrictive class: the attacker attacks at most $c$ targets, the defender can protect at most $k$ targets, where $c$ is a constant and $k$ is arbitrary; the defender's cost functions $C_d(\cdot)$ are additive. Then, the attacker's and defender's pure strategy spaces are given by $\mathcal{A}=\{A\in 2^{[n]}||A|\leq c\}, \mathcal{D}=\{D\in 2^{[n]}||D|\leq k\}$.
\begin{definition}
	A D-NASG is given by the tuple $(\mathcal{A}, \mathcal{D}, \mathcal{B}, \mathcal{C}^a$, $ \mathcal{C}^d)$, where the set of benefit function $\mathcal{B}=\{B(A)|A\in \mathcal{A}\}$, set of attacker's and defender's cost function $\mathcal{C}^a=\{C_a(A)|A\in \mathcal{A}\}$, $\mathcal{C}^d=\{C_d(i)|i\in [n]\}$.
\end{definition}
This assumption is motivated by the fact that both players have limited resources~\cite{kiekintveld2009computing} and they cannot cover any targets. Let $N_a=|\mathcal{A}|$ and $N_d=|\mathcal{D}|$ and our main result is the following theorem.
\begin{theorem}\label{thm:eqorandnasg}
There is a \emph{poly}$(n)$ time algorithm to compute the defender's mixed strategy in D-NASG, \textbf{if and only if} there is a \emph{poly}$(n)$ time algorithm to compute the defender oracle problem: for any $\mathbf{w}\in\mathbb{R}^{N_a}$,
\begin{equation}\label{eq:ddforacle}
	\mathbf{x}^*=\arg\max\limits_{\mathbf{x}\in H_d'}\mathbf{w}^T\mathbf{x},
\end{equation}
where the definition of $H_d'$ is given in (\ref{eq:polytope}).
\end{theorem}

\subsection{Reduction between D-NASG and Oracle Problem}

The reduction from D-NASG to DOP still follows our isomorphism and projection framework, and the main technical step is a partial decomposition of the payoff matrix. The reverse direction follows from a different path. First, let the payoff matrix of D-NASG be denoted by $\mathbf{M}^b\in \mathbb{R}^{|\mathcal{A}|\times |\mathcal{D}|}$, which is a sub-matrix of $\mathbf{M}^{\circ}$.
\begin{theorem}
The payoff matrix $\mathbf{M}^b\in\mathbb{R}^{N_a\times N_d}$ can be decomposed as
	\begin{equation}
	\mathbf{M}^b=\mathbf{I}\mathbf{M}^{\mathcal{A}}\mathbf{J}^T,
	\end{equation}	
	where the matrix $\mathbf{I}\in \mathbb{R}^{N_a\times N_a}$ and $\mathbf{J}\in \mathbb{R}^{N_d\times N_a}$ are binary matrices, and the matrix $\mathbf{M}^{\mathcal{A}}\in \mathbb{R}^{N_a\times N_a}$ contains one non-zero diagonal, non-zero row and non-zero column.
\end{theorem}
The detailed definition of each elements in matrix $\mathbf{I}, \mathbf{J}$ and $\mathbf{M}^{\mathcal{A}}$ can be seen in the supplemental material. Similarly, we have the affine transformation: $f'(\mathbf{p})=\mathbf{I}^T\mathbf{p}$, $g'(\mathbf{q})=\mathbf{J}^T\mathbf{q}$; transformed polytope: $\Delta_{N_a}^a$, $\Delta_{N_d}^d$; projected polytope: 
\begin{equation}
	H_a'=\Pi_{S}(\Delta_{N_a}^a), H_d'=\Pi_{S}(\Delta_{N_d}^d).\label{eq:polytope}
\end{equation}
Remark that in this case, the polytope $\Delta_{N_d}^d$ is not isomorphic with $\Delta_{N_d}$, but the correctness of our compact representation follows a similar proof of Theorem~\ref{thm:cpt}. Further, the compactly represented linear programming is expressed as
\begin{align}
\textbf{Compact-D-LP}\quad&\min\quad u\label{eq:d:compact}\\
& \begin{array}{r@{\quad}l@{}l@{\quad}l}
s.t. &\mathbf{v}^T\mathbf{M}^{\mathcal{A}}\mathbf{q'}\leq u\quad\forall\mathbf{v}\in I_a', \label{eq:d:payoffconstraint}\\
 &\mathbf{q'}\in H_d', 
\end{array} 
\end{align}
where $I_a'$ is the set of vertices of polytope $H_a'$. 
\begin{lemma}\label{lm:dorcpt}
	Separation problem for $H_d'$ and compact optimization problem (\ref{eq:d:compact}) reduces to each other in \emph{poly}$(n)$ time.
\end{lemma}
Considering the equivalence between the separation ($H_d'$) and optimization (DOP), we arrive at the reduction between DOP (\ref{eq:ddforacle}) and compact model (\ref{eq:d:compact}). The main technique in the reduction between D-NASG and compact optimization (\ref{eq:d:compact}) is: (i) for any arbitrary instance $\mathbf{M}^{\mathcal{A}}$ of compact optimization problem, we can construct the set of utility functions: $\mathcal{B}, \mathcal{C}^a$ in $O(2^cn)=O(n)$ time based on Lemma~\ref{lm:com}; (ii) vertex mapping algorithm from pure strategy to $H_d'$.
\begin{lemma}\label{lm:cptandorg}
	The min max problem of D-NASG and compact optimization (\ref{eq:d:compact}) reduces to each other in \emph{poly}$(n)$ time.
\end{lemma}
Lemma~\ref{lm:dorcpt} and Lemma~\ref{lm:cptandorg} together yield our desired result.

\subsection{What is the Defender Oracle Problem}

Through a series of reduction, we find that the NASG is essentially a defender oracle problem defined on a low-dimensional polytope $H_d'$, but the complicated form of polytope $H_d'$ still prevents us from uncovering the mystery of the NASG. Fortunately, based on the investigation of the geometric structure of the $H_d'$, we will prove that the DOP is indeed a problem of \emph{maximizing a pseudo-boolean function}.
\begin{theorem}
	The defender oracle problem is, for any vector $\mathbf{w}\in\mathbb{R}^{|S|}$, maximize a pseudo-boolean function under a cardinality constraint,
	\begin{equation}\label{eq:01opt}
	\max\limits_{\sum\limits_{i=1}^{n}\mathbf{x}_i\geq n-k,\mathbf{x}\in {\{0,1\}}^{n}}\left[\sum\limits_{V\in S}\mathbf{w}_{\sigma(V)}\left(\prod\limits_{\{i\}\in V} \mathbf{x}_i\right)\right].
\end{equation}
\end{theorem}
The complexity of (\ref{eq:01opt}) is dependent on the support set $S$. For example, in the simplest case, $S=[n]$, we can efficiently solve such a problem by summing all the positive elements of vector $\mathbf{w}$, which corresponds to the traditional additive security game. Instead, if $S=\{U\in 2^{[n]}||U|\leq 2\}$, then the oracle problem is a binary quadratic programming problem, which is known to be NP-hard. If $k=n$, the above problem will degenerate to the unconstrained optimization. This result builds a connection between the NASG and optimizating a \emph{pseudo-boolean function}, which enables us to design an efficient DOP solver or understand the complexity of NASG via analyzing the structure of the support set $S$ and using the results of combinatorial algorithm design.

\section{Application to the Network Security Domain}

In this section, we will apply our theoretical framework to an important domain, in which the security game occurs in a network. The following definition is motivated by the works~\cite{gueye2012towards,shakarian2014power}.
\begin{definition}
A network security game is given by the tuple ($G,T,\mathbf{F_a},c$), where $G=(V,E)$ with node set $V$, edge set $E$, $T$ is the network value function, $\mathbf{F_a}$ is the failure operator, $c$ is the maximum number of nodes attacker can choose and defender can protect any targets ($k=n$). 
\end{definition}

The network value function $T: G\rightarrow\mathbb{R}$ is a security measure assessing the utility of a network, and failure operator $\mathbf{F_a}: 2^G\rightarrow 2^G$ is to generate a new network via a specific failure mode after removing some nodes. For example, Shakarian et al.~\shortcite{shakarian2014power} adopts the number of connected load nodes as $T$, and edge cascading failure model as $\mathbf{F_a}$. The main result of our work is summarized as in TABLE~\ref{tab:one}.
\begin{table}[htb]%
\scriptsize
\centering
\vspace{-0.2in}
\caption{Solvability Status\label{tab:one}}{%
\begin{tabular}{|p{1.6cm}|p{1.5cm}|p{1.5cm}|c|}
\hline
\multicolumn{2}{|l|}{CASES} & SOLVABILITY& APPENDIX\\
\hline
\multicolumn{2}{|l|}{Additive benefit function} & poly($n$) & Trivial\\
\hline
\multicolumn{2}{|l|}{The separable support set $S$ with} & poly($n$) & Theorem 6,\\
\multicolumn{2}{|l|}{$\max_i|U_i|=\Theta(\log(n))$} &  & Corollary 1 \\
\hline
\multicolumn{2}{|l|}{Constant $c$, negative common utilities} & poly($n$)& Theorem 7,\\
\multicolumn{2}{|l|}{except for singleton set} &  & Corollary 2\\
\hline
Constant $c\geq 2$ & \multicolumn{2}{|c|}{NP-hard and efficient approximation} & Last Section\\
\hline
\end{tabular}}
\vspace{-0.1in}
\end{table}%

The first polynomial solvable class is trivial. The basic idea of second solvable class is to show that when $S$ is separable with size of largest component equal to $\Theta(\log(n))$, the DOP is separable and can be solved in poly$(n)$ time via an enumerating algorithm. Here the ``separable'' is defined as, $S= \bigcup_{i=1}^{m} S_i$ such that $A_i\cap A_j=\emptyset, \forall A_i\in S_i, A_j\in S_j, i\neq j$, the component is defined as $U_i=\cup_{U\in S_i} U$. In the third solvable class, we will show that DOP under such condition is indeed a submodular minimization problem, which can be solved in poly$(n)$ time. One may wonder why these special cases are interesting. In fact, they correspond to the following applications. 
\begin{itemize}
	\item The second class can be applied in a sparse network. For example, if the size of largest connected component of $G$ is $\Theta(\log(n))$, $S$ will satisfy the condition of second solvable class (Corollary 1 in supplemental material).
	\item The third class can be applied to a dense network where the most nodes are adjacent. For example, if $c=2$, attacking any two nodes will lead to the superposition of the failure effect, resulting in negative common utilities. 
	\item Another application of third class: in cybersecurity, the sensor network often exhibits a tree topology. The game is such that the attacker attempts to invade some nodes to destroy the connectedness of the network and the IT manager is required to deploy the anti-virus software in some nodes. We can show that this game satisfies the condition of the third class. (Corollary 2 in supplemental material)
\end{itemize}

\begin{algorithm}[t]
\renewcommand{\algorithmicrequire}{ \textbf{Input:}} %Use Input in the format of Algorithm  
\renewcommand{\algorithmicensure}{ \textbf{Output:}} %UseOutput in the format of Algorithm 
\caption{Separable Approximation}   
\label{alg2}  
\begin{algorithmic}  
%\REQUIRE Network $G,T,\mathbf{F},c$  
%\ENSURE Common utility $\mathcal{B}^c$ and $\{S_i\}_{i=1}^{m}$.
\STATE Calculate benefit $B(U)=T(G)-T(\mathbf{F_a}(G\backslash U)), \forall U\in\mathcal{A}$; 
\FOR{each $U\in\mathcal{A}$}
\STATE Calculate $B^c(U)=\sum_{W\subseteq U}(-1)^{|U\backslash W|}B(W)$;
\STATE \textbf{if} $|B^c(U)|\leq \epsilon_c$ \textbf{ then} $\tilde{B}^c(U)=0$;
\ENDFOR
\STATE Create support set $S$ and let $(S_1,\ldots,S_m)\leftarrow$ disjoint ($S$);
\end{algorithmic}
\end{algorithm}

For the general network (not sparse, not dense or not a tree), the problem is clearly NP-hard and there exists two problems: (1) can we still compactly represent the game if $c$ is large, i.e., $|S|=\text{poly}(n)$? (2) can we efficiently solve such a game? To tackle these problems, we propose a novel separable approximation framework (Algorithm~\ref{alg2}), which can guarantee the approximation error of the original problem, instead of the DOP. One crucial observation is that 

\emph{the common utility in the realistic network is well concentrated around zero.}

Then we can let most of the common utility functions equal to $0$ based on a given threshold $\epsilon_c$. Considering Lemma~\ref{lm:com} and a result of Lipton et al.~\shortcite{lipton2003playing}, the approximation error of the game value can be guaranteed by $\epsilon=2^{c+1}\epsilon_c$. As we will show via simulations in the supplemental material, this process will greatly reduce the size of the support set $S$ and lead to a separable structure of $S$. The resulting complexity of solving NASG is poly$(n)O(2^{\max_i|U_i|})$. For example, in most networks including Erd$\ddot{o}$s-Renyi, scale-free network and a $39-$nodes Italian communication network, only $10\%$ approximation error can reduce the complexity term ${\max_i|U_i|}$ to the order of $\Theta(\log(n))$. Therefore, using our theoretical framework, we can \textbf{approximately and compactly represent a realistic network security game and solve it in poly$(n)$ time with high accuracy}.

\section{Conclusion}

In this paper, we examined the security game under non-additive utility functions and a structured strategy space (uniform matroid). We showed that the size of compact representation is dependent on the number of non-additive strategies, and NASG is essentially the problem of \emph{maximizing a pseudo-boolean function}. Here the non-additive strategies refer to those have non-zero common utilities. Compared with previous results, this work greatly extends the polynomial solvable class, provides an understanding of the complexity properties, and partly answers the question proposed by Xu~\shortcite{xu2016mysteries} in zero-sum, uniform matroid scenario.

For future directions, we plan to investigate (i) the relationship between the Oracle problem and the NASG when the defender has a non-structured strategy space; (ii) how to efficiently compute the defender's mixed strategy when attacker and defender have different benefit functions.

\bibliographystyle{aaai}
\bibliography{Sinong}

\end{document}